\documentclass[
aps,
prl,
tightenlines,
superscriptaddress,
nofootinbib,
floatfix,
showpacs,
groupedaddress,
preprintnumbers,
nofootinbib,
twocolumn,
amsmath,
amsfonts,
amssymb]
{revtex4}
\usepackage{graphicx,
longtable,
color
}

\usepackage[a4paper, portrait, margin = 0.5 in]{geometry}

\usepackage{array}
\newcolumntype{L}[1]{>{\raggedright\let\newline\\\arraybackslash\hspace{0pt}}m{#1}}
\newcolumntype{C}[1]{>{\centering\let\newline\\\arraybackslash\hspace{0pt}}m{#1}}
\newcolumntype{R}[1]{>{\raggedleft\let\newline\\\arraybackslash\hspace{0pt}}m{#1}}
\newcommand{\la}{\lambda}

\newcommand{\wt}{\widetilde}
\newcommand{\ov}{\overline}
\usepackage{cancel}
 \usepackage[normalem]{ulem}
\usepackage{color}

\begin{document}

\preprint{HRI-RECAPP-2016-012, IP/BBSR/2016-9}

\title{LHC diphoton excess in a left-right symmetric model with minimal dark matter}

\author{                Sanjib Kumar Agarwalla}\email{sanjib@iopb.res.in}
\affiliation{           Institute of Physics, Sachivalaya Marg, Sainik School Post, Bhubaneswar 751005, India}

\author{                Kirtiman Ghosh}\email{kirti.gh@gmail.com}
\affiliation{           Department of Physics and Astrophysics, University of Delhi, Delhi 110007, India}

\author{                Ayon Patra}\email{ayon@okstate.edu}
\affiliation{            Institute of Convergence Fundamental Studies, Seoul National University of Science and Technology, Seoul 139-743, Korea}
\affiliation{            Regional Centre for Accelerator-based Particle Physics, Harish-Chandra Research Institute, Jhunsi, Allahabad - 211019, India${^\star}$}


\begin{abstract}

We construct a model containing a viable dark matter candidate, in the framework of left-right symmetry, 
which can explain both width and cross-section of the observed 750 GeV diphoton excess at the LHC.
We introduce a fermion quintuplet whose neutral component can be a possible dark matter candidate, 
whereas the charged components enhance the loop induced coupling of a 750 GeV singlet scalar 
with a pair of photons. We study the photo-production of the singlet scalar and its various decay modes, 
successfully addressing both the ATLAS and the CMS diphoton excess.

\end{abstract}


\maketitle

At the intensity frontier, the ongoing Large Hadron Collider (LHC) experiment is expected to shed light 
on new physics scenarios to break apart our long-held understanding of the Standard Model (SM) of 
particle physics, and the recent tantalizing hint of a new resonance, giving rise to an excess in the diphoton 
invariant mass around 750 GeV, may be the first glimpse of that 
new physics~\cite{ATLAS-Diphoton-1,CMS:2015dxe,ATLAS-Diphoton-2,CMS:2016owr,Aaboud:2016tru,Khachatryan:2016hje}.
This observed excess of diphoton events\footnote{For a recent review on the main experimental, phenomenological, 
and theoretical issues related to the 750 GeV diphoton excess, see Ref.~\cite{Strumia:2016wys}.}
at the early 13 TeV run of the LHC experiment~\cite{Aaboud:2016tru,Khachatryan:2016hje} 
could be explained as production of a new massive spin-0 or spin-2 resonance followed by its decay into a pair of photons.
The most significant deviation from the SM background prediction was observed at $M_{\gamma \gamma} 
\sim 750~(760)$ GeV by the ATLAS (CMS) collaboration. The 13 TeV ATLAS data with 3.2 fb$^{-1}$ 
integrated luminosity indicates towards an excess in favor of a large width ($\Gamma \sim$ 45 GeV $\sim$ 0.06 M)
resonance, with a local significance of $3.9\sigma$ ($ 3.6\sigma$) in searches optimized for spin-0 (spin-2)
hypothesis. For CMS collaboration with $\sqrt s$ = 13 TeV and 3.3 fb$^{-1}$ data, the local significance is 
maximized to 2.8$\sigma$ (2.9$\sigma$) assuming spin-0 (spin-2) hypothesis for a resonance with 
$\Gamma \sim$ 10 GeV $\sim$ 0.014 M. The best-fit values for the effective signal strength 
(namely, production cross-section times branching ratio to $\gamma\gamma$) of a 750 GeV 
resonance are ($5.5\pm 1.5$) fb and ($4.8 \pm 2.1$) fb for ATLAS and CMS, respectively with 
13 TeV data~\cite{Strumia:2016wys}. 

To explain this excess, a plethora of beyond the SM scenarios have been put forward in the literature 
(see Ref.~\cite{Strumia:2016wys} and the references therein). The obvious (and also maximally 
studied)
explanation is a new spin-0 resonance (scalar or pseudoscalar) for which coupling to a pair of 
gluons/photons appears only at the loop level\footnote{Randall-Sundrum type graviton couples to a pair of gluons at tree-level and hence, 
could explain the excess as spin-2 resonance. However, in this case, it is non-trivial to suppress 
its decay to a pair of top-quarks or leptons since neither of them have been 
seen~\cite{Arun:2015ubr,Falkowski:2016glr,Csaki:2016kqr,Hewett:2016omf}.}. Assuming gluon-gluon fusion to be the dominant 
production mechanism for the resonance, the observed signal cross-section requires additional 
colored and charged states. They need to have large coupling to the resonance to enhance
its loop induced coupling to a pair of gluons/photons, which ultimately increases the production
cross-section and diphoton decay width. Initial state gluons being colored objects are more prone 
to emit hadronic radiation (initial state radiation, ISR) and thus, give rise to hadronic activity in the central 
region. Therefore, if gluon-gluon fusion is the production mechanism, the diphoton signal events are expected 
to be accompanied by a few ISR jets, which should have observable features in kinematical distributions 
like number of jets ($n$--jets) distribution, transverse momentum ($p_T$) distribution of diphotons {\em etc.}
The ATLAS collaboration with limited luminosity has recently reported jet multiplicity distribution and 
diphoton $p_T$ distribution in the diphoton excess region and its sideband~\cite{Aaboud:2016tru}.
In view of these distributions, it was already pointed out in Ref.~\cite{Dalchenko:2016dfa,Harland-Lang:2016vzm} 
that gluon-gluon fusion is slightly disfavored.

The other possible production mechanism for a 750 GeV scalar is the photo-production {\em i.e.,} 
production via photon-photon fusion~\cite{Fichet:2015vvy,Csaki:2015vek,Csaki:2016raa,Abel:2016pyc,Fichet:2016pvq,Salvio:2016hnf,Barrie:2016ntq,Barrie:2016ndh,Ghosh:2016lnu}. Initial state photon being color singlet, the hadronic activity 
from ISR would be suppressed in photo-production, resulting in fewer central jets which is consistent with recent ATLAS 
observation~\cite{Aaboud:2016tru}. The photo-production cross-section of a resonance 
($R$) at the LHC can be estimated from its decay width into a pair of photons and 
parton distribution function of photon. The diphoton signal cross-section at LHC is given 
by~\cite{Csaki:2016raa}:
\begin{equation}
\sigma(R\to \gamma\gamma) = \sigma_0 \left(\frac{\Gamma_{\rm TOT}}{{\rm GeV}}\right){\rm Br}^2(R\to \gamma \gamma),
\label{eq:cross}
\end{equation} 
where $\Gamma_{\rm TOT}$ is the total decay width of resonance $R$ and $\sigma_0$ 
is estimated\footnote{$\sigma_0$ gets contribution from fully inelastic ($\sim 63\%$), 
partially inelastic ($\sim 33\%$), and elastic ($\sim 4\%$) proton-proton scattering.} 
to be around 240 fb~\cite{Csaki:2016raa} at the LHC with $\sqrt s=13$ TeV.
Eq.~\ref{eq:cross} shows that large total decay width of resonance $R$ and sizable (few \%) 
branching ratio to diphoton could explain the cross-section and width of the observed 
diphoton excess. 

The loop induced coupling of a scalar with a pair of photons is quadratically proportional to the 
charge of the particle inside the loop. In this work, we consider a minimal model~\cite{Cirelli:2005uq,Ko:2015uma} 
for dark matter (DM) in the framework of $SU(3)_C\times SU(2)_L\times SU(2)_R\times U(1)_{B-L}$ 
gauge symmetry~\cite{Mohapatra:1974hk,Senjanovic:1975rk}, where $B$ and $L$ are baryon and 
lepton numbers respectively. Apart from the usual chiral fermions, scalar doublet, and bi-doublet required 
to break left-right (LR) symmetry, the particle spectrum includes an additional $SU(2)_R$ vector-like fermion 
quintuplet to accommodate a DM candidate (the neutral component of the quintuplet) and a singlet scalar
to explain the LHC diphoton excess. The loop-induced diphoton decay width of the singlet scalar can be 
significantly enhanced due to multi-charged (depending on the $B$--$L$ charges) quintuplet fermions 
in the loop and hence, could potentially explain the LHC diphoton excess in totality. 
Before going into the details of diphoton signal cross-section and width of the signal distribution, 
we present our model briefly.

The matter content of this model along with their gauge quantum numbers in the framework of 
$G_{3221} \equiv SU(3)_c \times SU(2)_L \times SU(2)_R \times U(1)_{B-L}$ gauge symmetry 
are summarized in the following:
\begin{eqnarray}
&Q_L& \left(3,2, 1, \frac13 \right ) = \!\left (\begin{array}{c}
u\\ d \end{array} \right )_L,
Q_R\left ( 3,1, 2, \frac13
\right )\!=\!\left (\begin{array}{c}
u\\d \end{array} \right )_R,\nonumber \\
&l_L&\left ( 1,2, 1, -1 \right )=\left (\begin{array}{c}
\nu\\ e\end{array}\right )_L,
l_R\left ( 1,1, 2, -1 \right )=\left (\begin{array}{c}
\nu \\ e \end{array}\right )_R,\nonumber \\
&N& \left( 1,1,1,0 \right),~\chi(1,1,5,4) = ( \chi^{4+},\chi^{3+},\chi^{2+},\chi^{+},\chi^0)^T,\nonumber
\end{eqnarray}
where the Majorana fermion $N$ is required for neutrino mass generation through 
inverse seesaw mechanism~\cite{Mohapatra:1986bd,Barr:2005ss}, and $\chi$ is the 
vector-like fermion quintuplet under $SU(2)_R$ with a $B$--$L$ charge of 4. 
The minimal set of scalar multiplets\footnote{For a study of various scalar sectors 
in the supersymmetric left-right scenario, see~\cite{Babu:2014vba}.}~\cite{Babu:1987kp,Ma:1986we} 
required for the spontaneous breaking of $G_{3221}$ to the SM, and then to $U(1)_{\rm EM}$ 
are as follows:
\begin{eqnarray}
H_R(1,1,2,1)&=&\left (\begin{array}{c}
H_R^+ \\H_R^0 \end{array} \right ),
\Phi(1,2,2,0)={\left (\begin{array}{cc}
\phi^{0}_1 & \phi^{+}_{2} \\ \phi^{-}_{1} & \phi^{0}_{2} \end{array} \right)}.~~~
\label{scalar}
\end{eqnarray}
The $SU(2)_R \times U(1)_{B-L}$ breaks down to $U(1)_Y$ as the neutral component of
$H_R$ acquires a non-zero vacuum expectation value (VEV) denoted as $\left< H^0_R \right> = v_R$.
On the other hand, VEVs of $\Phi$ namely,  $\left< \phi_{1}^0 \right> = v_1$ and $\left< \phi^0_{2} \right> = v_{2}$, 
are responsible for the electroweak (EW) symmetry breaking and for generation of the SM fermion masses and mixings.
The electric charge $Q$ is given as: $Q=T^3_L+T^3_R+Q_{(B-L)}/2$. In addition to the scalar doublet and bi-doublet in 
Eq.~\ref{scalar}, we also introduce a singlet scalar $S(1,1,1,0)$ to explain the LHC diphoton excess.

The structure of $SU(2)_L\times SU(2)_R \times U(1)_{B-L}$ breaking to $U(1)_{\rm EM}$ introduces 
mixings between the gauge bosons of $SU(2)_L$, $SU(2)_R$, and $U(1)_{B-L}$ resulting in four 
massive ($W_R,~Z_R$, and the SM $W$ and $Z$-boson) and one massless (the SM photon) gauge bosons:
\begin{eqnarray}
M^2_{W_R}&=&\frac{1}{2}g_R^2\left( v_R^2+v^2 \right),\nonumber\\
M^2_{Z_R} &=& \frac{1}{2}\left(g_R^2+g^2_{B-L}\right) \left[ v_R^2+\frac{g_R^2 v^2}{\left(g_R^2+g^2_{B-L}\right)} \right],
\end{eqnarray}
where $v^2=v_1^2+v_2^2$ is the EW VEV $\sim$ 174 GeV and $g_R = g_L = 0.653$.
The left-handed $W$ and $Z$ boson masses are the same as in the SM 
with $g_Y = (g_R g_{B-L})/(g_R^2+g_{B-L}^2)^{1/2}$. The relevant couplings of 
the gauge bosons with $\chi$ are given as:
\begin{eqnarray}
L \supset &-&g_Y s_W Q \overline{\chi} Z^\mu \gamma_\mu \chi + e Q \overline{\chi} A^\mu \gamma_\mu \chi \nonumber \\
&+& \sqrt{g_R^2-g_Y^2}\left[Q - \frac{g_R^2 Q_{B-L}}{2 \left(g_R^2-g_Y^2\right)} \right]\overline{\chi} Z_R^\mu \gamma_\mu \chi,
\end{eqnarray}
where $s_W = \sin{\theta_W}$ with $\theta_W$ being the Weinberg angle.

The quark and charged lepton masses are generated from the following 
Yukawa Lagrangian:
\begin{eqnarray}
\mathcal{L}_Y&=&Y^q\overline{Q}_{L}\Phi Q_{R}+\widetilde{Y}^q\overline{Q}_{L}\widetilde{\Phi}Q_{R}+Y^l\overline{l}_{L}\Phi l_{R}+\widetilde{Y}^l\overline{l}_{L}\widetilde{\Phi}l_{R} \nonumber \\
&+& f_{R}\overline l_{R} \widetilde H_R N + \frac{{\mu_N}}{2} N N + H.C.
\end{eqnarray}
where $Y$ and $f$ are the Yukawa couplings and $\widetilde{\Phi}=\tau_2\Phi^\ast\tau_2,\widetilde H_{R} = i \tau_2 H^\ast_{R}$.
The quark and charged lepton masses in this model would then be given as:
\begin{eqnarray}
M_{u} &=& Y^q v_1+\wt{Y}^q v_2,M_d=Y^q v_2+\wt{Y}^q v_1, \nonumber \\
M_l &=& Y^l v_2+\wt{Y}^l v_1,~~~~
\label{eq:fermass}
\end{eqnarray}
while the neutrino masses are generated through the inverse seesaw mechanism. 
For simplicity, we will choose a large $\tan{\beta}~(=v_1/v_2)$ limit which requires 
$Y_{33}^q \sim 1$ to explain the top mass while $\wt{Y}_{33}^q < 10^{-2}$.

The scalar sector consists of a bidoublet field, an $SU(2)_R$ doublet field, and 
a real singlet. The most general scalar potential involving these fields is given by:
\begin{eqnarray}
&&V_{H} \supset - \mu_1^2~ \text{Tr}\left[\Phi^{\dagger} \Phi \right] - \mu_2^2~\text{Tr} \left[\widetilde \Phi \Phi^{\dagger} + H.C. \right] - \mu_R^2 H_R^\dagger H_R \nonumber \\
&-& \frac{\mu_S^2}{2} S^2  + \lambda_1 \left[ \text{Tr} \left(\Phi^{\dagger} \Phi\right) \right]^2+ \lambda_4 \text{Tr} \left(\Phi^{\dagger} \Phi \right) \text{Tr} \left[\widetilde \Phi \Phi^{\dagger} + H.C. \right] \nonumber \\
&+&\lambda_2 \left[ \left\{ \text{Tr} \left( \wt \Phi \Phi^{\dagger}\right) \right\}^2 +H.C. \right]+ \lambda_3 \text{Tr} \left( \wt \Phi \Phi^{\dagger}\right) \text{Tr} \left( {\wt \Phi}^{\dagger} \Phi\right) \nonumber \\
&+& \alpha_3 \mu_3 S \text{Tr}\left(\Phi^{\dagger} \Phi \right) + \alpha_4 \mu_4 S \text{Tr} \left[\widetilde \Phi \Phi^{\dagger} + H.C. \right] \nonumber\\
&+& \alpha_5 \mu_5 S H_R^\dagger H_R + \frac{\beta_1}{2} S^2 \text{Tr}\left(\Phi^{\dagger} \Phi \right) +  \frac{\beta_2}{2} S^2 \text{Tr} \left[\widetilde \Phi \Phi^{\dagger} + H.C. \right]  \nonumber\\
&+& \frac{\beta_3}{2} S^2 H_R^\dagger H_R + \rho_1 H_R^\dagger H_R  \text{Tr}\left[\Phi^{\dagger} \Phi \right] + \rho_3 H_R^\dagger \Phi \Phi^\dagger H_R \nonumber\\
&+&  \rho_2 H_R^\dagger H_R \text{Tr} \left[\widetilde \Phi \Phi^{\dagger} + H.C. \right] + \rho_4 H_R^\dagger \Phi^\dagger \Phi H_R \nonumber \\
&+& \lambda_R \left( H_R^\dagger H_R \right) ^2 + A_S \mu_S S^3 + \frac{\lambda_S}{4} S^4.
\end{eqnarray} 
The Higgs spectrum of this model consists of four CP-even scalars, one CP-odd pseudoscalar, 
and one charged Higgs boson. Two CP-odd states, and two charged states are eaten up by the 
four {massive} gauge bosons. After diagonalizing the mass-squared matrix, one can easily obtain 
a light SM-like 125 GeV Higgs boson {denoted by $h$}, consisting almost entirely of the real part 
of $\phi_1^0$ field. We also get a 750 GeV scalar denoted by $H_1$, consisting of almost purely 
the singlet $S$ with negligible mixing with the others. Two very heavy states $H_2$ and $H_3$ 
with masses of the order of $v_R$ consisting of real part of $\phi_2^0$ and $H_R^0$ states 
are also present in the spectrum. {The heavy states are required to be heavier than 15 TeV 
in order to suppress flavor changing neutral currents~\cite{Ecker:1983uh,Mohapatra:1983ae,Pospelov:1996fq,Zhang:2007da,Maiezza:2010ic}. 
This can be easily satisfied in our model by choosing a high value ($>$ 10 TeV) for the 
right-handed symmetry breaking scale, $v_R$. Our model can accommodate the above mentioned
scalar mass spectrum for a wide range of parameter choice. A typical set of parameters is given 
in Tab~{\ref{tab:one}} as a benchmark point (BP). The mass of the pseudo-scalar $A_0$ and the 
charged Higgs boson $H^{\pm}$ are also proportional to $v_R$, and hence, are heavier than
15 TeV.
\begin{table}[h!]
\centering
\begin{tabular}{||C{1.1cm}|C{1.8cm}|C{5.5cm}||}
\hline
Particle & Mass & Composition \\ \hline
$h$& 125 GeV & $0.999 \phi_1^0+0.029 \phi_2^0 - 0.010 S-0.003 H_R^0 $ \\ \hline
$H_1$ & 751 GeV & $0.010 \phi_1^0 + 0.0002 \phi_2^0 + 0.999 S  - 0.001 H_R^0$ \\ \hline
$H_2$ & 18.36 TeV & $0.020 \phi_1^0 - 0.617 \phi_2^0 +0.001 S + 0.787 H_R^0 $ \\ \hline
$H_3$ & 18.43 TeV & $0.021 \phi_1^0 - 0.787 \phi_2^0 - 0.001 S - 0.617 H_R^0 $ \\ \hline
\end{tabular}
\caption{Scalar mass eigenstates for $\la_1$ = 0.15, $\la_2$ = -0.35, $\la_3$ = 0.831, 
$\la_4$ = 0.01, $\alpha_3$ = 0.14, $\alpha_4$ = 0.1, $\mu_3$ = 174 GeV, $\mu_4$ = 174 GeV, 
$\beta_1$ = 0.1, $\beta_2$ = 0.1, $\beta_3$ = 0.004, $\rho_1$ = 0.2, $\rho_2$ = 0.2, 
$\rho_3$ = 1, $\rho_4$ = 1, $\mu_S$ = 340 ${\text{GeV}}$, $v_1$ = 173.9 GeV, 
$v_2$ = 5 GeV.}  
\label{tab:one}  
\end{table}

The original motivation for introducing the vector-like quintuplet fermion $\chi(1,1,5,4)$ 
was to obtain a candidate for DM~\cite{Cirelli:2005uq}. The components of quintuplet 
are mass degenerate at tree-level, but radiative corrections remove this degeneracy. 
The mass splitting due to quantum corrections is given by,
\begin{eqnarray}
M_{\chi^Q}-M_{\chi^0}&=&\frac{g_R^2}{(4\pi)^2}M {\bigg{[}} Q\left(Q-Q_{B-L}\right)f(r_{W_R}) \nonumber \\
&-& Q \left(\frac{g_Y^2}{g_{B-L}^2}Q-Q_{B-L} \right)f(r_{Z_R})\nonumber \\
&-& \left. \frac{g_Y^2}{g_R^2}Q^2\left\{s_W^2 f(r_{Z}) + c_W^2 f(r_{\gamma}) \right\} \right],
\label{eq:mass}
\end{eqnarray}
where $r_X$ = $m_X/M$ and $f(r) \equiv 2 \int_0^1 dx (1+x) \text{log} \left[ x^2 + (1-x)r^2 \right]$. 
The masses of the charged components of the quintuplet get positive contribution from the 
radiative corrections and hence, $\chi^0$ becomes the lightest among the quintuplet fermions. 
Being weakly interacting and lightest member of the quintuplet, $\chi^0$ could be a good candidate 
for dark matter. The stability of $\chi^0$ is ensured by its gauge quantum numbers. 
Being a part of the quintuplet, $\chi^0$ can decay to the SM particles via interactions with
dimension-6 or higher operators and thus, its decay width is suppressed atleast by a factor 
of $1/\Lambda^2$. For a TeV scale $\chi^0$, the decay width via dimension-6 operator is of the 
order of $M^3/\Lambda^2$ which corresponds to a lifetime greater than the age of the universe 
for $\Lambda \gtrsim 10^{14}$ GeV.

In this work, we consider the scalar $H_1$, which is dominantly singlet, as a candidate for the 
750 GeV diphoton resonance observed at the LHC. Being singlet, $H_1$ has Yukawa interaction 
only with the quintuplet fermion. It has also couplings (arising from the scalar potential) with a pair 
of Higgs and other heavy scalars. The relevant interactions for collider phenomenology 
related to $H_1$ are given by:
\begin{eqnarray}
L_S \supset \alpha_3 \mu_3 H_1 h h +  \lambda_{\chi} \ov{\chi} \chi H_1,
\label{coup_h1}
\end{eqnarray}
where $\lambda_{\chi}$ is the Yukawa coupling, $\alpha_3$ is a dimensionless parameter, 
and $\mu_3$ is a parameter with a dimension of mass. Therefore, apart from its mass ($m_{H_1}$), 
the collider phenomenology of $H_1$ depends on $\mu_3$ which we choose to be equal 
to the EW VEV (174 GeV), $\alpha_3$ and $\lambda_{\chi}$. 

\begin{figure}[t!]
\vspace*{-0.3 cm}
\includegraphics[angle=0,width=0.5\textwidth]{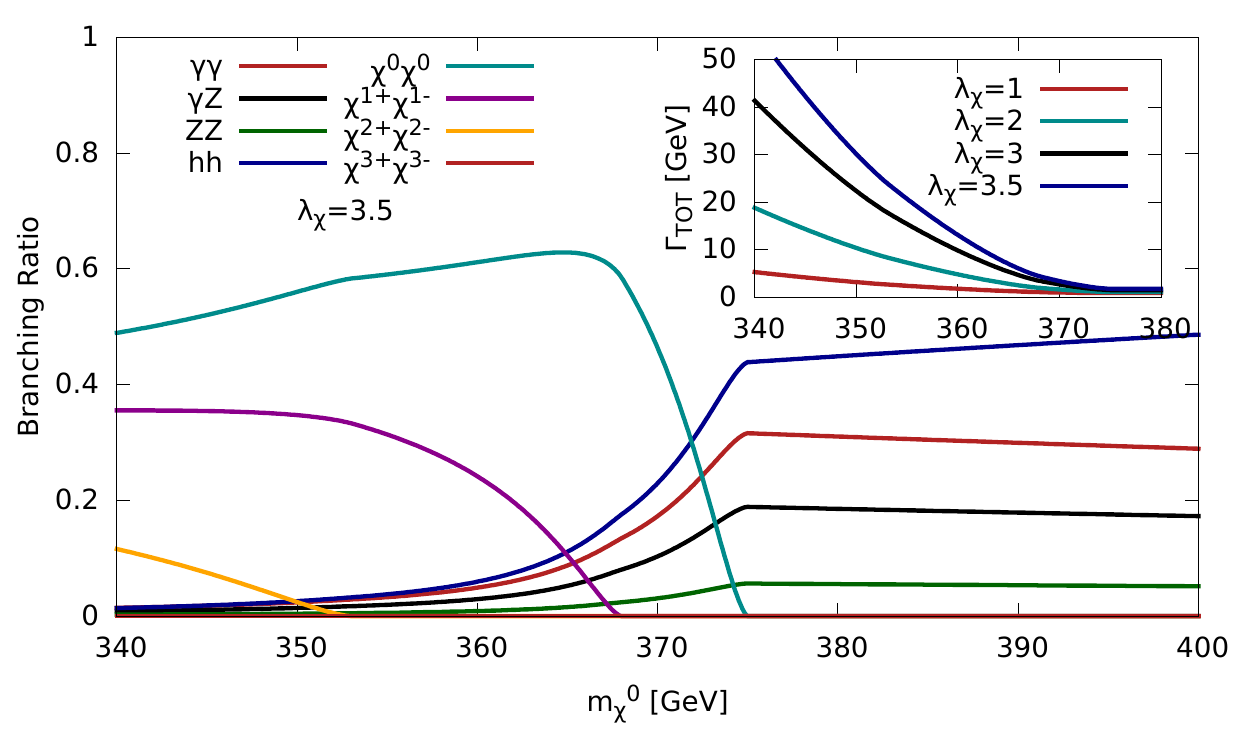}
\vspace*{-0.5 cm}
\caption{Branching ratios of $H_1$ into different decay modes 
as a function of mass of $\chi^0$ for a fixed value of $\lambda_\chi = 3.5$.
Inset shows the total decay width of $H_1$ as a function of $m_{\chi^0}$ 
for different values of $\lambda_{\chi}$.}
\label{branching}
\end{figure}

The dominant decay modes of $H_1$ are its tree level decay into a pair of SM Higgs or a pair 
of quintuplet fermions (if kinematically possible). $H_1$ can also decay into a pair of photons or pair 
of $Z$-bosons or $Z\gamma$ pairs via loops involving quintuplet fermions. The loop induced 
decay into photons is quartically proportional to the electric charge of the loop fermion.
Therefore, in presence of multi-charge quintuplet fermions in the loop, the decay width of 
$H_1$ into a pair of photons gets enhanced by orders of magnitude in our model. 
This motivates us to study the diphoton signature of $H_1$ produced via photon-photon fusion 
at the LHC. It was shown in Eq.~\ref{eq:cross} that the diphoton signal cross-section depends on 
the total decay width ($\Gamma_{\rm TOT}$) and the branching ratio into 
$\gamma\gamma$ of $H_1$. The decay widths for different decay channels are given by,
\begin{eqnarray}
\Gamma_{hh}&=&\frac{\alpha_3^2 \mu_3^2}{8 \pi m_{H_1}}\left(1-\frac{4 m_{h}^2}{m_{H_1}^2}\right)^{\frac{1}{2}},\nonumber\\
\Gamma_{\chi_i \bar \chi_i}&=&\frac{\lambda_\chi^2}{8\pi}m_{H_1}\left(1-\frac{4 m_{\chi_i}^2}{m_{H_1}^2}\right)^{\frac{3}{2}},\nonumber\\
\Gamma_{\gamma\gamma}&=&\frac{\alpha^2 m_{H_1}^3 \lambda_\chi^2}{256 \pi^3}\left|\sum_{\chi_i} \frac{Q^2_{\chi_i}}{m_{\chi_i}}A_{\frac{1}{2}}\left(\frac{m_{H_1}^2}{4m_{\chi_i}^2}\right)\right|^2,\nonumber\\
\Gamma_{Z\gamma}&=&\frac{\alpha^2 m_{H_1}^3 \lambda_\chi^2}{128 \pi^3}{\rm tan}^2\theta_W\left|\sum_{\chi_i} \frac{Q^2_{\chi_i}}{m_{\chi_i}}A_{\frac{1}{2}}\left(\frac{m_{H_1}^2}{4m_{\chi_i}^2}\right)\right|^2,\nonumber\\
\Gamma_{ZZ}&=&\frac{\alpha^2 m_{H_1}^3 \lambda_\chi^2}{256 \pi^3}{\rm tan}^4\theta_W\left|\sum_{\chi_i} \frac{Q^2_{\chi_i}}{m_{\chi_i}}A_{\frac{1}{2}}\left(\frac{m_{H_1}^2}{4m_{\chi_i}^2}\right)\right|^2,~~~~
\label{decay_width}
\end{eqnarray}
where $\chi_i \subset \{\chi^{++++},~\chi^{+++},~\chi^{++},~\chi^{+}~{\rm and}~\chi^{0}\}$. 
$m_{\chi_i}$ and $Q_{\chi_i}$ are the mass and charge of the corresponding $\chi_i$ respectively. 
The loop function $A_{1/2}(x)$ is given by, $A_{1/2}(x)=2x^{-2}\left[x+(x-1)f(x)\right]$, 
where, $f(x)=-[{\rm ln}\{(1+\sqrt{1-x})/(1-\sqrt{1-x})\}-i\pi]^2/4$ for $x >1$ 
and $f(x)={\rm arcsin}^2{\sqrt x}$ for $x \le 1$. Eqs.~\ref{decay_width} shows that the total decay width 
and hence, the branching ratios crucially depend on the masses of the quintuplet fermions. 
In order to calculate the masses of quintuplet fermions which depend on the masses of 
$W_R$ and $Z_R$ (see Eq.~\ref{eq:mass}), we assume $v_R=13.0$ TeV which corresponds 
to $m_{W_R}=6.0$ TeV and $m_{Z_R}=7.2$ TeV. In Fig.~\ref{branching}, we show the 
branching ratios of $H_1$ into different decay modes as a function of $\chi^0$ mass 
denoted by $m_{\chi^0}$ for a fixed value of $\lambda_\chi = 3.5$. The inset of Fig.~\ref{branching} 
depicts the total decay width as a function of $m_{\chi^0}$ for different values of $\lambda_{\chi}$. 
Fig.~\ref{branching} clearly demonstrates that $H_1$ dominantly decays into a pair of quintuplet 
fermions if the decay is kinematically possible and the total decay width is quite large 
in this case (see the inset). If the decay to $\chi_i \bar\chi_i$ is forbidden then the dominant decay 
mode is a pair of the SM Higgs bosons where the decay to diphoton being the second dominant mode.

\begin{figure}[t!]
\vspace*{-0.85 cm}
\hspace*{-0.80 cm}
\includegraphics[height = 6.5 cm, width = 10.0 cm]{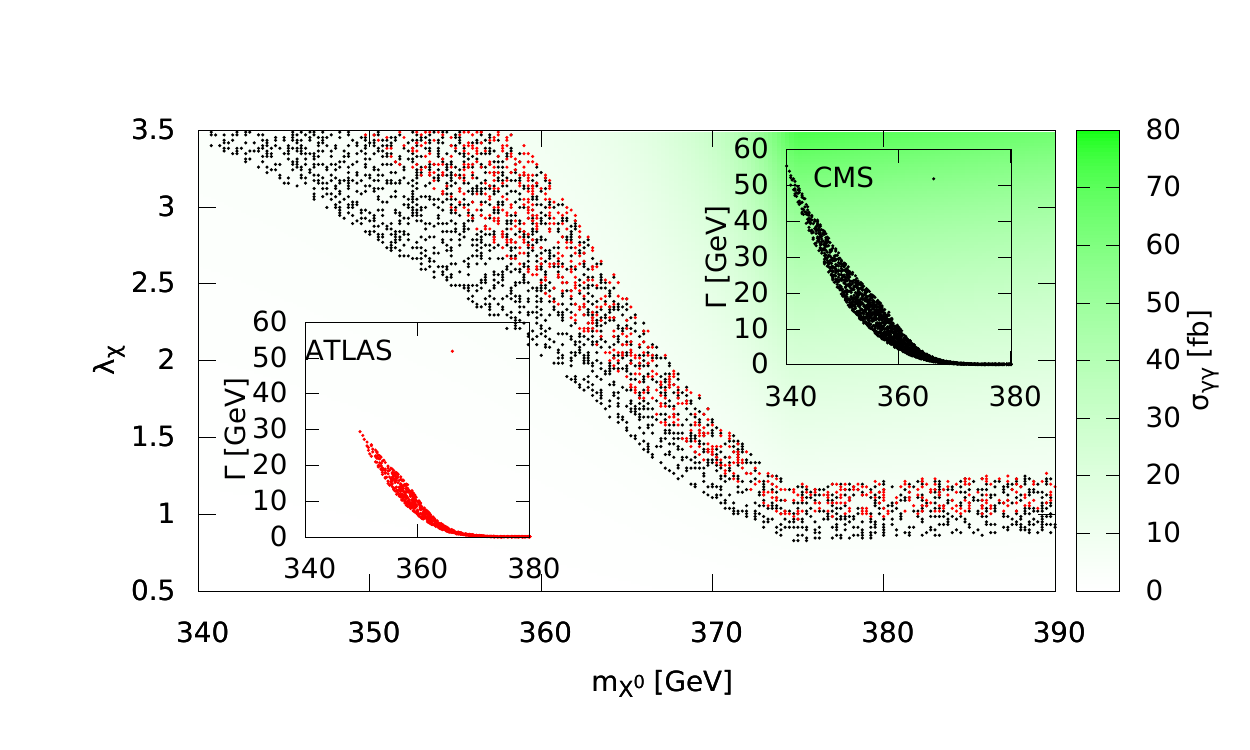}
\vspace*{-0.75 cm}
\caption{Cross-section for the 750 GeV resonance decaying into diphoton is presented 
by color gradient as a function of DM-mass ($m_{\chi^0}$) and Yukawa coupling 
($\lambda_{\chi}$). The red and black dotted regions correspond to the cross-sections 
consistent with the observed diphoton excess at the ATLAS and CMS detectors, respectively. 
The total decay width of the 750 GeV resonance for red and black dotted points are 
shown in the inset.}
\label{cross}
\end{figure}

The large decay width and significant branching ratio into a pair of photons in 
Fig.~\ref{branching} for $m_{\chi_i} < m_{H_1}/2$ indicates towards the possibility 
of explaining the cross-section and width of LHC diphoton excess via the photo-production 
of $H_1$ and followed by its decay into $\gamma \gamma$. In Table~\ref{BPs}, 
we have presented our numerical results for three different benchmark points defined 
by $\lambda_\chi$ and $m_{\chi^0}$. Table~\ref{BPs} contains quintuplet mass spectrum, 
relevant decay widths (including the total decay width), and diphoton signal cross-section 
at the 13 TeV LHC for a $H_1$ with 750 GeV mass. It also shows that the chosen BPs are 
consistent (both cross-section and width) with the LHC observed diphoton excess. 
In Fig.~\ref{cross}, we have shown our model prediction for the diphoton cross-section 
by color gradient for a scan over the parameter space defined by $m_{\chi^0}$ (along x-axis) 
and $\lambda_\chi$ (along y-axis). We also indicate the region of parameter space consistent with 
the ATLAS and CMS measured excess by red and black dots, respectively. 
The total decay width corresponding to the red and black dots are depicted in the insets of 
Fig.~\ref{cross}. There is a large region of parameter space in our model that gives a diphoton 
signature which is consistent with the LHC observed excess in totality 
({\em i.e.,} both the cross-section and width of the excess). 

Finally, we arrive at a discussion on the present bounds (from collider as well as DM experiments), 
and future confirmatory tests for our model. The explanation for the cross-section and width of LHC 
diphoton excess requires the DM mass to be $\sim m_{H_1}/2$. Therefore, the DM annihilation 
cross-section is enhanced due to the resonant contribution from $H_1$ and hence, the upper bound 
on the DM relic density from WMAP/PLANK data~\cite{Komatsu:2010fb,Ade:2015xua} can be easily 
satisfied in this particular part of parameter space. However, the large $B$--$L$ charge of the 
quintuplet results in an enhanced DM-nucleon scattering cross-section and hence, stringent 
constraints arise from direct DM detection experiments. $\chi^0$ interacts with nucleon via 
exchange of a $Z_R$-boson and thus, the DM-nucleon cross-section is suppressed by 
$1/m_{Z_R}^{4}$. In Fig.~\ref{DM}, we give the $\chi^0$-proton and $\chi^0$-neutron 
scattering cross-section as a function of $m_{Z_R}$. The experimental bound on DM-nucleon 
scattering cross-section depends on DM-mass. The shaded region (in the inset of Fig.~\ref{DM}) 
in $m_{DM}$--$m_{Z_R}$ plane is consistent with the LUX~\cite{Akerib:2013tjd} upper bound on 
the DM-nucleon scattering cross-section. It also clearly shows that the DM-mass $\sim$
375 GeV is allowed for $m_{Z_R}$ $\sim$ 7 TeV.

\begin{figure}[t]
\vspace*{0.2 cm}
\includegraphics[width=0.48\textwidth]{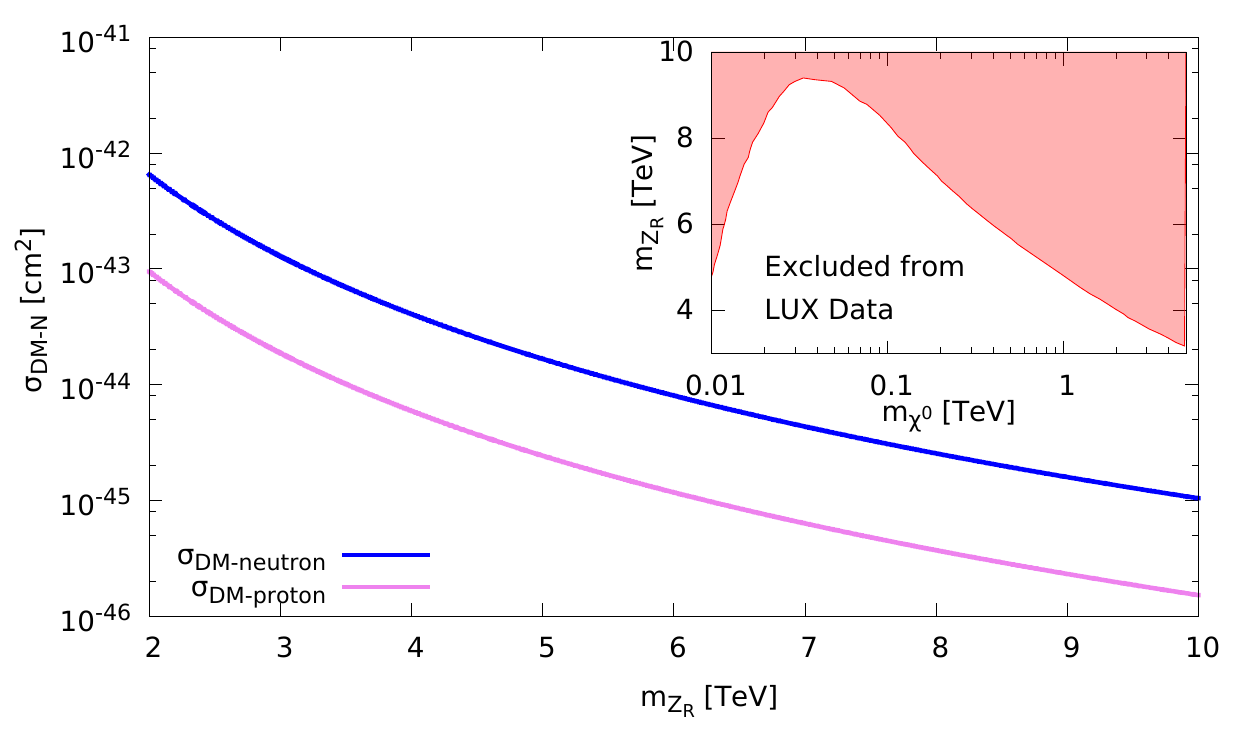}
\vspace*{-0.75 cm}
\caption{$\chi^0$-proton and $\chi^0$-neutron scattering cross-section as a function 
of $m_{Z_R}$. In the inset, the shaded region in $m_{DM}$--$m_{Z_R}$ plane is 
consistent with the LUX~\cite{Akerib:2013tjd} upper bound on DM-nucleon scattering 
cross-section.}
\vspace{-0.2cm}
\label{DM}
\end{figure}

At the LHC, the pair-production of quintuplet fermions take place via quark-antiquark 
($s$-channel $Z/\gamma^*$-exchange) or photon-photon ($t$-channel) initial states. 
A quintuplet fermion decays into next lighter quintuplet fermion in association with a 
lepton-neutrino pair or quark-antiquark pair via tree-level 3-body decay involving an
off-shell $W_R^*$. Though the cross-sections are suppressed, the pair-production 
of charged quintuplet fermions give rise to spectacular multi-lepton 
(4 same sign leptons (SSL), 3-SSL, 2-SSL, 8, 7, 6, 5 lepton {\em etc.}) signatures at the LHC. 
However, due to small mass splitting between the quintuplet fermions 
(see Table~\ref{BPs}), the resulting leptons are usually soft and fall below the minimum 
transverse momentum threshold required at the LHC for most of the events. A dedicated 
collider study is required to probe the spectacular leptonic signatures of our model.

To summarize, we have explained the the cross-section and width of both ATLAS
and CMS diphoton excess by postulating a 750 GeV singlet scalar in the framework 
of a left-right model which also gives a viable candidate for DM in the form of the neutral 
component of an $SU(2)_R$ vector-like fermion quintuplet. The loop induced coupling 
of the 750 GeV scalar with $\gamma\gamma$ is enhanced by multi-charged quintuplet 
fermions. We have studied the photo-production of this scalar followed by its decay 
to $\gamma\gamma$ at the 13 TeV LHC. We have also discussed the bounds from 
the direct DM detection experiment such as LUX as well as future collider signatures 
of our proposed model.    

\onecolumngrid

\begin{table}[h!]
\begin{ruledtabular}
\begin{tabular}{||c|c||c|c|c|c||c|c|c||c|c|c|c||c||c||}
$\lambda_\chi$ & $m_{\chi^0}$ & $m_{\chi^{1+}}$& $m_{\chi^{2+}}$& $m_{\chi^{3+}}$& $m_{\chi^{4+}}$ & $\sigma_{DM-n} \times $ & $\sigma_{DM-p} \times $ & $\sigma_{LUX} \times$ & $\Gamma_{\gamma\gamma}$& $\Gamma_{Z\gamma}$& $\Gamma_{ZZ}$& $\Gamma_{hh}$& $\Gamma_{TOT}$ & $\sigma_{\gamma \gamma}$ \\
&GeV & GeV & GeV & GeV & GeV & $10^{45}$[cm$^2$] & $10^{46}$[cm$^2$] & $ 10^{45}$[cm$^2$] & GeV & GeV & GeV & GeV & GeV & fb \\\hline\hline
3.51  & 349.0 &357.2 & 374.1 & 399.8 & 434.1 & & & & 0.73 & 0.43 & 0.13 & 0.78 & 31.9 &4.0\\
2.52 & 355.0 & 363.3 & 380.5 & 406.4 & 441.2 &3.9 &5.7 &4.2 & 0.35 & 0.21 & 0.06 & 0.78 & 11.1 &2.7\\
1.99  & 374.0 & 382.7 & 400.5 & 427.4 & 463.4 & & & & 0.18 & 0.11 & 0.03 & 0.78 & 1.15 &7.1\\
\end{tabular}
\end{ruledtabular}
\caption{Quintuplet fermions mass spectrum, DM-neutron, DM-proton scattering cross-sections, and 
corresponding LUX~\cite{Akerib:2013tjd} bound, decay width of 750 GeV singlet scalar into 
$\gamma\gamma$, $Z\gamma$, $ZZ$, $hh$ as well as total decay width and diphoton cross-section 
at 13 TeV LHC are shown for three benchmark points defined by $\lambda_{\chi}$ and DM-mass ($m_{\chi^0}$). 
We consider $g_R/g_L=$ 1.0 and $m_{Z_R}=7.2$ TeV.}
\label{BPs}
\end{table}

\newpage
\twocolumngrid


\section*{Acknowledgments}

S.K.A. is supported by the DST/INSPIRE Research Grant [IFA-PH-12],
Department of Science \& Technology, India. The work of K.G. is supported 
by an Inspire Faculty Fellowship from the Government of India, 
at the University of Delhi. A.P. is currently visiting 
Harish-Chandra Research Institute (HRI) and would like to thank the 
Regional Centre for Accelerator-based Particle Physics, HRI for hospitality 
during the visit, when part of this work was done. We would also like to thank 
Prof. Kaladi Babu for useful discussions.

\bibliographystyle{apsrev-title}
\bibliography{collider}

\end{document}